\documentclass[aps,twocolumn,showpacs,pra,preprintnumbers,amsmath,amssymb]{revtex4}

\usepackage[dvipdfmx]{graphicx}
\usepackage{dcolumn}
\usepackage{bm}

\usepackage[dvipdfmx]{color}
\usepackage{ulem}

\begin{document}

\preprint{APS/123-QED}

\title{
Temperature evolution of spin dynamics in two- and three-dimensional Kitaev models:\\ 
Influence of fluctuating gauge fluxes 
}

\author{Junki Yoshitake$^1$, Joji Nasu$^2$, and Yukitoshi Motome$^1$}
\affiliation{%
$^1$Department of Applied Physics, University of Tokyo, Bunkyo, Tokyo 113-8656, Japan\\
$^2$Department of Physics, Tokyo Institute of Technology, Meguro, Tokyo 152-8551, Japan
}%




\date{\today}

\begin{abstract}
The long-sought quantum spin liquid is a quantum-entangled magnetic state leading to the fractionalization of spin degrees of freedom. 
Quasiparticles emergent from the fractionalization affect not only the ground state properties but also thermodynamic behavior in a peculiar manner. 
We here investigate how the spin dynamics evolves from the high-temperature paramagnet to the quantum spin liquid ground state, for the Kitaev spin model describing the fractionalization into itinerant matter fermions and localized $Z_2$ gauge fluxes. 
Beyond the previous study [J. Yoshitake, J. Nasu, and Y. Motome, Phys. Rev. Lett. \textbf{117}, 157203 (2016)], in which the mean-field nature of the cluster dynamical mean-field theory prevented us from studying low-temperature properties, we develop a numerical technique by applying the continuous-time quantum Monte Carlo (CTQMC) method to statistical samples generated by the quantum Monte Carlo (QMC) method in a Majorana fermion representation. 
This QMC+CTQMC method is fully unbiased and enables us to investigate the low-temperature spin dynamics dominated by thermally excited gauge fluxes, including the unconventional phase transition caused by gauge flux loops in three dimensions, which was unreachable by the previous methods. 
We apply this technique to the Kitaev model in both two and three dimensions. 
Our results clearly distinguish two cases: while the dynamics changes smoothly through the crossover in the two-dimensional honeycomb case, it exhibits singular behaviors at the phase transition in the three-dimensional hyperhoneycomb case. 
We show that the low-temperature spin dynamics is a sensitive probe for thermally fluctuating gauge fluxes that behave very differently between two and three dimensions. 
\end{abstract}

\maketitle

\section{Introduction}
\label{sec:intro}

The quantum spin liquid (QSL) is an exotic state of matter in insulating magnets showing no magnetic order down to zero temperature ($T$)~\cite{Anderson1973,Balents2010}. 
It is not characterized by any conventional order parameter, but known to exhibit topological quantum entanglement resulting in fractionalization of the fundamental spin degrees of freedom~\cite{Wen1991,Misguich2011}. 
This is purely quantum mechanical nature arising in strongly correlated many-body systems, as seen in fractional charges by the fractional quantum Hall effect~\cite{Tsui1982,Stormer1999}. 
Although the spin fractionalization has attracted great attention for identifying the QSL in candidate materials, the unambiguous detection remains largely elusive~\cite{Balents2010,Yamashita2008,Yamashita2009,Yamashita2010}.

The Kitaev spin model, originally introduced on a two-dimensional (2D) honeycomb lattice~\cite{Kitaev2006}, has generated a new trend in the study of QSLs. 
This is because of the following virtues of this model. 
First of all, the model is exactly soluble in the ground state, and the exact ground state is a QSL. 
The exact solution is obtained by representing the spin operators by Majorana fermion operators, which simultaneously provides canonical formulation of the fractionalization: the elementary spin excitations are described by itinerant matter fermions and localized $Z_2$ gauge fluxes, both of which are composed of the Majorana fermions. 
Furthermore, the model can be extended to any tri-coordinate lattices with preserving the solubility, even in three dimensions (3D)~\cite{Mandal2009,Hermanns2015,O'Brien2016}. 
Last but not least, the bond-dependent anisotropic interaction in this model has a realization in some magnetic materials with strong spin-orbit coupling~\cite{Jackeli2009}. 
All these features have accelerated the combined studies between theory and experiment for realization and identification of Kitaev QSLs~\cite{Nussinov2015,Trebst_preprint}.

Among many consequences of the spin fractionalization unveiled by the recent studies of the Kitaev model is thermal fractionalization, i.e., thermodynamic signatures originating from different energy scales of the fractionalized quasiparticles~\cite{Nasu2015}. 
The thermal fractionalization manifests itself in, for instance, two peaks in the specific heat at $T=T_L$ and $T_H$ ($T_L \ll T_H \sim J$, where $J$ is the dominant Kitaev coupling) and successive entropy release by a half of $\log 2$ around these temperatures. 
Besides, the spin dynamics is also of importance for experimental identification of the fractionalization.
In the previous studies~\cite{Yoshitake2016,Yoshitake_preprint}, the authors calculated dynamical quantities for the 2D Kitaev model, developing the cluster extension of the dynamical mean-field theory (CDMFT) in a Majorana fermion representation and combining it with the continuous-time quantum Monte Carlo (CTQMC) method. 
The CDMFT+CTQMC study revealed an interesting aspect of the fractionalization: dichotomy between static and dynamical spin correlations. 
This was shown by the significant $T$ evolution of the magnetic susceptibility $\chi$, the NMR relaxation rate $1/T_1$, and the dynamical spin structure factor $S(\mathbf{q},\omega)$ in the $T$ regime below $T_H$ where the static spin correlations saturate and almost $T$ independent.

Despite the successful calculations of dynamical properties, the applicable $T$ range of the CDMFT+CTQMC method is limited: the method does not give reasonable results at very low $T$. 
This is due to the occurrence of phase transition at $T\sim T_L$ as an artifact of the mean-field approximation in the CDMFT. 
Moreover, the CDMFT+CTQMC method is not suitable for the Kitaev model on 3D lattices by the following reasons. 
One is that a larger cluster is necessary in the CDMFT, as the unit cell, or more strictly speaking, the smallest loop of lattice sites, for which the conserved $Z_2$ gauge flux is defined, becomes larger for 3D than 2D in general. 
Another reason is that the 3D extensions of the Kitaev model may cause a phase transition, which might be hard to capture by the CDMFT. 
For instance, the Kitaev model on a 3D hyperhoneycomb lattice exhibits an unconventional  phase transition triggered by proliferation of loops composed of thermally excited gauge fluxes~\cite{Nasu2014b,Nasu2014}. 
The cluster approximation in the CDMFT is not suitable to describe such a topological transition characterized by global quantities beyond the cluster. 
An alternative method is desired to study the spin dynamics, including the low-$T$ behavior. 

Besides such a theoretical demand, it is crucial to clarify the spin dynamics of the Kitaev model in the whole $T$ range also from the experimental point of view. 
Recently, many candidates have been explored in both quasi-2D and 3D materials~\cite{Singh2010,Singh2012,Plumb2014,Takayama2015,Modic2014}. 
Some indications of the fractionalization were observed, for instance, in the specific heat~\cite{Mehlawat2017}, magnetic Raman scattering~\cite{Sandilands2015,Glamazda2016}, inelastic neutron scattering~\cite{Banerjee2016,Banerjee_preprint,Do_preprint}, and thermal transport~\cite{Hirobe_preprint,Leahy2017}. 
However, such indications are for rather high-$T$ features, corresponding to the theoretical predictions around and below $T_H$ associated with itinerant matter fermions~\cite{Nasu2015,Nasu2016,Yoshitake2016,Yoshitake_preprint,Nasu_preprint}. 
It is highly desired to experimentally capture another indications dominated by thermally excited gauge fluxes at lower $T$. 
Although all the candidate materials exhibit a magnetic order at low $T$, several efforts have been made for suppressing the order, e.g., by external pressure~\cite{Takayama2015,Breznay_preprint}, magnetic field~\cite{Ruiz_preprint,Hentrich_preprint,Wolter_preprint}, and chemical substitution~\cite{Lampen-Kelley_preprint}. 
Given such an upsurge of interest, it is highly important to clarify the dynamical behavior of the 2D and 3D Kitaev models down to the lowest $T$. 

In this paper, we propose a new numerical method which overcomes the problems in the previous CDMFT+CTQMC method. 
We here adopt the quantum Monte Carlo (QMC) method, instead of the CDMFT, for generating statistical samples used in the CTQMC calculations. 
The QMC method is also formulated on the basis of a Majorana fermion representation, which has been used to compute thermodynamic properties in a series of previous studies for the Kitaev models on several tri-coordinate lattices~\cite{Nasu2015,Nasu2014b,Nasu2015b,Nasu2016,Nasu_preprint}. 
Thus, the new combined method, which we call the QMC+CTQMC method, provides a versatile technique, free from biased approximation. 
We demonstrate that the method is applicable in a wider $T$ range, including the low-$T$ region below $T\sim T_L$, which was not accessible by the previous CDMFT+CTQMC method. 
In the 2D honeycomb case, comparing the data of $\chi$ and $1/T_1$ by the CDMFT+CTQMC and QMC+CTQMC methods, we show that although the former works quite well above the fictitious critical temperature, only the latter can give reasonable results at lower $T$. 
In the 3D hyperhoneycomb case, we present the QMC+CTQMC results for $\chi$, $1/T_1$, and $S(\mathbf{q},\omega)$. 
From the comparison between the 2D and 3D results, we clarify the signatures arising from the difference of the system dimension. 
While everything changes smoothly through the crossover at $T=T_L$ in the 2D honeycomb case, the dynamical quantities exhibit singular behaviors in the 3D hyperhoneycomb case at the phase transition caused by the topological nature of excited gauge flux loops. 
Thus, the QMC+CTQMC is applicable to the unconventional phase transition in 3D, which is not accessible by the CDMFT+CTQMC method. 
Our results show that the dynamical properties at low $T$ depend substantially on the system dimension, despite almost dimension-independent behavior of the static spin correlations. 
This is the low-$T$ aspect of the dichotomy between static and dynamical spin correlations, which was found in the intermediate $T$ region in the previous study. 

The structure of this paper is as follows. 
In Sec.~\ref{sec:model_method}, we introduce the Kitaev model and its Majorana fermion representation. 
We also present the details of the QMC+CTQMC method. 
In Sec.~\ref{sec:results}, we present the QMC+CTQMC results for the 2D and 3D cases in Sec.~\ref{subsec:2D} and \ref{subsec:3D}, respectively.
Finally, Sec.~\ref{sec:summary} is devoted to the summary.

\section{Model and method}
\label{sec:model_method}

\begin{figure}[htp]
    \includegraphics[width=\columnwidth,clip]{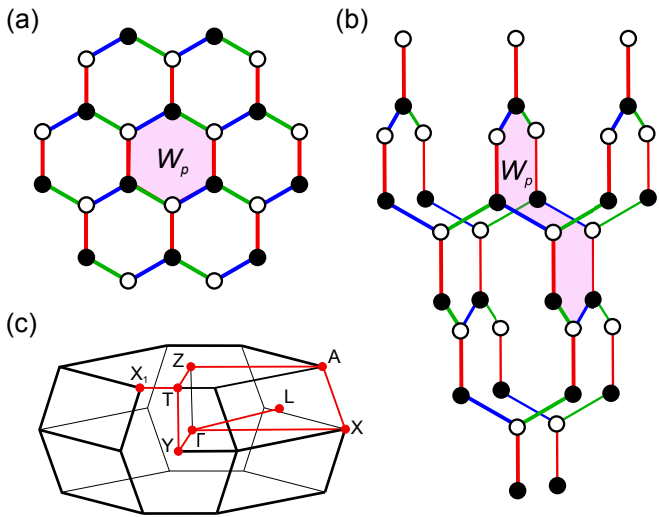}
    \caption{ \label{fig:fig1} 
    Schematic picture of the Kitaev model on (a) the 2D honeycomb lattice and (b) the 3D hyperhoneycomb lattice. 
    The blue, green, and red bonds represent the $x$, $y$, and $z$ bonds in Eq.~(\ref{eq:H_Kitaev}), respectively. 
    The black and white circles denote the sites $j$ and $j'$ in Eq.~(\ref{eq:H_Majorana}), respectively. 
    (c) displays the first Brillouin zones for the 3D case, in which the red lines indicate the symmetric lines used for the plot in Fig.~\ref{fig:fig6}.
    }
\end{figure}

In this study, we consider the Kitaev model on a 2D honeycomb lattice [Fig.~\ref{fig:fig1}(a)] and 3D hyperhoneycomb lattice [Fig.~\ref{fig:fig1}(b)], whose Hamiltonian is given in the common form~\cite{Kitaev2006,Mandal2009}
\begin{align}
\mathcal{H} = 
-\sum_p J_p \sum_{\langle j,j'\rangle_p} S_j^p S_{j'}^p.
\label{eq:H_Kitaev}
\end{align}
Here, $p=x,y,z$ represents one of the three different types of bonds on the tri-coordinate lattices, and $\langle j,j'\rangle_p$ denotes a set of neighboring sites $j,j'$ on the $p$ bonds; see Figs.~\ref{fig:fig1}(a) and \ref{fig:fig1}(b). 
$S_j^p$ represents the $p$ component of quantum spin $S=1/2$ at site $j$, and $J_p$ is the coupling constant for the $p$ bond.

A mathematically faithful representation of the Hamiltonian in Eq.~(\ref{eq:H_Kitaev}) is obtained by applying the Jordan-Wigner transformation along the chains composed of the $x$ and $y$ bonds~\cite{Chen2007,Feng2007,Chen2008}:
\begin{align}
\mathcal{H}=
i\frac{J_x}{4} \sum_{(j,j')_x} c_{j'} c_j
- i\frac{J_y}{4} \sum_{(j,j')_y} c_j c_{j'}
- i\frac{J_z}{4} \sum_{(j,j')_z} \eta_r c_j c_{j'},
\label{eq:H_Majorana}
\end{align}
where $c_j$ and $\bar{c}_j$ are two types of Majorana fermion operators at site $j$; 
$\eta_r = i\bar{c}_j \bar{c}_{j'}$ is defined on each $z$ bond connecting sites $j$ and $j'$. 
The sum over $(j,j')_p$ is taken for the neighboring sites $j$ and $j'$ colored by black and white, respectively, in Figs.~\ref{fig:fig1}(a) and \ref{fig:fig1}(b).
The bond variable $\eta_r$ commutes with the Hamiltonian as well as other $\eta_{r'}$, and $\eta_r^2=1$; hence, $\eta_r$ is a conserved $Z_2$ variable taking $\pm 1$. 
The ground state is exactly obtained as the state with all $\eta_r=+1$ for both honeycomb and hyperhoneycomb cases. 
The exact ground state is shown to be a QSL, both gapless and gapped depending on the ratios between the coupling constants $J_p$~\cite{Kitaev2006}. 
The elementary excitations are also exactly described by the operators $\{c_j \}$ and $\{\eta_r \}$. 
In this Majorana fermion representation, therefore, the original spin operators $\{\mathbf{S}_j \}$ are fractionalized into $\{c_j \}$, which describe itinerant Majorana fermions called matter fermions, and $\{\eta_r \}$, which are the localized $Z_2$ variables. 

The $Z_2$ variables $\{\eta_r \}$ are related with the $Z_2$ gauge fluxes $W_p$ discussed in the original paper by Kitaev~\cite{Kitaev2006}. 
The gauge flux is also a conserved $Z_2$ quantity defined for each elementary plaquette [a hexagon in the 2D honeycomb case and a ten-site plaquette in the 3D hyperhoneycomb case; see Figs.~\ref{fig:fig1}(a) and \ref{fig:fig1}(b)]: it is defined by the product of $\eta_r$ belonging to the plaquette $p$, as $W_p = \prod_{r\in p} \eta_r$. 
The ground state with all $\eta_r=+1$ corresponds to the state with all $W_p=+1$, which is called the flux-free state. 
At nonzero $T$, the gauge fluxes are thermally excited from the flux-free state by flipping $W_p$. 

In the previous study, the authors have developed the CDMFT+CTQMC method for calculating the finite-$T$ spin dynamics of the Kitaev model in Eq.~(\ref{eq:H_Kitaev}), by using the Majorana representation in Eq.~(\ref{eq:H_Majorana})~\cite{Yoshitake2016,Yoshitake_preprint}. 
In this method, we generate the configurations of the $Z_2$ variables $\{\eta_r \}$ by the CDMFT, and compute the imaginary-time spin correlations by applying the CTQMC calculations to each configuration. 
The combined method successfully delivers precise data for the dynamical properties in a wide $T$ range. 
A problem in the CDMFT+CTQMC method is that the cluster approximation in the CDMFT part leads to a fictitious phase transition at $T=\tilde{T}_c$ by ordering of $\{\eta_r \}$. 
In the 2D Kitaev model on the honeycomb lattice, there is no phase transition at a nonzero $T$ and only two crossovers occur at very different $T$ scales, $T_L$ and $T_H$~\cite{Nasu2015}. 
In the isotropic case with $J_x=J_y=J_z=J=\pm 1$, $T_L\simeq 0.012$ and $T_H\simeq 0.375$; the fictitious $\tilde{T}_c\simeq 0.014$ is slightly higher than $T_L$. 
Meanwhile, in the 3D case on the hyperhoneycomb lattice, the model exhibits a phase transition at $T=T_c$ ($T_c\simeq 0.0039$ for the isotropic case), but it is not due to the ordering of $\{\eta_r \}$: the transition is caused by global objects, i.e., closed loops composed of thermally excited gauge fluxes $W_p$~\cite{Nasu2014b}. 
Thus, the phase transition by ordering of $\{\eta_r \}$ in the CDMFT is an artifact arising from the mean-field nature. 
Because of this problem, the CDMFT+CTQMC method is not applicable to the very low-$T$ region around and below $T_L$ in 2D and $T_c$ in 3D~\footnote{At sufficient low $T$, where almost all $\eta_r=+1$, the CDMFT+CTQMC method reproduces well the quantum spin liquid nature.}. 

In order to solve this problem, instead of the CDMFT, we here adopt the real-space QMC simulation, which has been used to calculate static quantities in the previous studies~\cite{Nasu2015,Nasu2014b,Nasu2015b,Nasu2016,Nasu_preprint}. 
Using the QMC simulation, we generate statistical samples of the configuration of localized $Z_2$ variables $\{\eta_r \}$, for which the dynamical spin correlations are computed by the CTQMC simulation. 
In this case, we can study much larger system sizes than the clusters used in the CDMFT, which enables us to systematically investigate the low-$T$ dynamical properties including the unconventional phase transition in 3D without biased approximation. 
We call this new combined technique the QMC+CTQMC method.

In Sec.~\ref{sec:results}, we compute the dynamical properties for the isotropic case with $J_x=J_y=J_z=J=\pm 1$ by the QMC+CTQMC method; $J=+1$ corresponds to the ferromagnetic (FM) case, while $J=-1$ the antiferromagnetic (AFM) case. 
All the static quantities, such as the specific heat, behave in the same manner for the FM and AFM cases, and hence, the crossover and phase transition temperatures are common to the two cases. 
The configurations of $\{\eta_r \}$ are generated by the QMC calculations under the same conditions with the previous studies~\cite{Nasu2014b,Nasu2015}. 
Note that the QMC simulation is done for finite-size clusters with the open boundary condition, at least, in one direction. 
For each configuration, we perform the CTQMC calculations for the $z$ bonds, sufficiently far from the open boundaries. 
Typically, we select $60$ ($16$-$18$) bonds in the 2D (3D) case near the central region of each cluster (away from the open boundary), and average the results over the bonds. 
In each CTQMC calculation, we typically perform $5\times 10^3$ measurements at every $20$ MC steps, after $10^5$ MC steps for initial relaxation.
To obtain the dynamical quantities as functions of the real frequency from the imaginary-time spin correlations, we perform the maximum entropy method (MEM) under the same conditions with the previous CDMFT+CTQMC study~\cite{Yoshitake_preprint}; we use the Legendre polynomial up to $100$th order for $T\leq 0.006$, while we expand up to $50$th order for higher $T$ as well as for the 2D case.

\section{Results}
\label{sec:results}

\subsection{2D honeycomb}
\label{subsec:2D}

\begin{figure}[htp]
    \includegraphics[width=.88\columnwidth,clip]{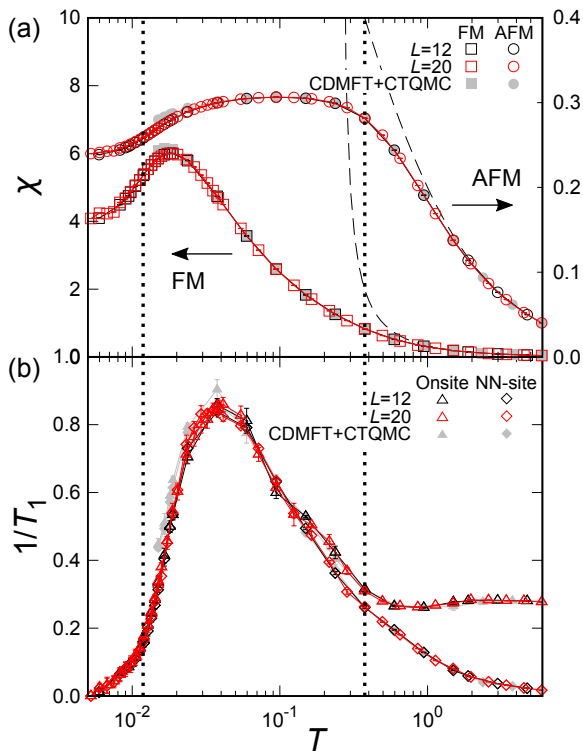}
    \caption{ \label{fig:fig2} 
    QMC+CTQMC results for the 2D honeycomb Kitaev model with isotropic $J_p$: (a) the magnetic susceptibility $\chi$ and (b) the NMR relaxation rate $1/T_1$. 
    In (a), FM and AFM denote the ferromagnetic case with $J_p=J=1$ and the antiferromagnetic case with $J_p=J=-1$, respectively. 
    While the onsite component of $1/T_1$ is common to the FM and AFM cases, the NN-site component for the AFM case is obtained by changing the sign of the FM data plotted in (b). 
    For comparison, we plot the CDMFT+CTQMC results in Ref.~\cite{Yoshitake2016} by gray symbols.
    The vertical dotted lines indicate $T_L\simeq 0.012$ and $T_H\simeq 0.375$ (see Ref.~\cite{Nasu2015}).
    In (a), the dashed curves represent the Curie-Weiss behaviors, $\chi_{\rm CW} = 1/(4T-J)$. 
    }
\end{figure}

First, we show the results for the 2D case on the honeycomb lattice.
Figure~\ref{fig:fig2} displays the QMC+CTQMC results for the magnetic susceptibility $\chi$ and the NMR relaxation rate $1/T_1$. 
$\chi$ is calculated from the imaginary-time spin correlations, without using the MEM, as
\begin{align}
\chi = \frac{1}{N}
\sum_{j,j'} \int_0^\beta d\tau\langle S^z_j (\tau) S^z_{j'}\rangle, 
\label{eq:chi}
\end{align}
where $N$ is the system size and $\beta=1/T$ is the inverse temperature (we set the Boltzmann constant $k_B=1$ and the reduced Planck constant $\hbar=1$). 
On the other hand, we compute $1/T_1$ by~\cite{Yoshitake_preprint} 
\begin{align}
1/T_1 = S^x_{j,j}(\omega = 0) + S^y_{j,j}(\omega = 0),
\label{eq:1/T_1}
\end{align}
for the onsite component and
\begin{align}
1/T_1 = S^x_{j,j'}(\omega = 0) + S^y_{j,j''}(\omega = 0),
\label{eq:1/T_1_nnsite}
\end{align}
for the nearest-neighbor(NN)-site component separately, 
where $S^p_{j,j'}(\omega)$ is the spin correlations as a function of the real frequency $\omega$ obtained by the MEM from $\langle S^p_j (\tau) S^p_{j'}\rangle$. 
Here, $j'$ and $j''$ are the sites neighboring to site $j$ on the $x$ and $y$ bonds, respectively. 
Note that both $\chi$ and $1/T_1$ are isotropic in spin space for the current isotropic case with $J_x=J_y=J_z$ on the honeycomb lattice. 

As shown in Figs.~\ref{fig:fig2}(a) and \ref{fig:fig2}(b), the results for different system sizes $L=12$ and $20$ agree with each other ($N=2L^2$), indicating that the QMC+CTQMC results well converge with respect to the system size. 
In the figures, the previous CDMFT+CTQMC results are also plotted by gray symbols for comparison~\cite{Yoshitake2016}. 
In the CDMFT+CTQMC method, as mentioned above, the cluster mean-field approximation leads to a fictitious phase transition at $\tilde{T}_c\simeq 0.014$, and hence, we plot the data above $\tilde{T}_c$. 
We find that the QMC+CTQMC results well agree with the CDMFT+CTQMC ones for $T\gtrsim \tilde{T}_c$, which supports the validity of the latter for $T\gtrsim \tilde{T}_c$. 
While such validity was claimed for the static quantities in the previous studies~\cite{Yoshitake2016,Yoshitake_preprint}, the present results demonstrate it explicitly for the dynamical quantities. 

The present QMC+CTQMC method enables us to study the low-$T$ region around and below the low-$T$ crossover temperature $T_L\simeq 0.012$, beyond $\tilde{T}_c$ in the CDMFT+CTQMC result. 
$T_L$ is the temperature where the localized $Z_2$ gauge fluxes $W_p$ begin to be frozen into the flux-free state while decreasing $T$~\cite{Nasu2015}. 
Thus, our results show how the dynamical properties are affected by thermally excited gauge fluxes. 
Figure~\ref{fig:fig2}(a) indicates that, while decreasing $T$ around $T_L$, $\chi$ decreases slightly and changes the curvature from upward to downward convex, for both the FM and AFM cases. 
While further decreasing $T$, $\chi$ appears to converge to a nonzero value, as expected for the system which does not conserve the $z$ component of total spin. 
The asymptotic value is almost one order of magnitude larger for the FM case than the AFM case. 
On the other hand, as shown in Fig.~\ref{fig:fig2}(b), $1/T_1$ decreases below the peak slightly above $T_L$ as partly seen in the CDMFT+CTQMC results~\cite{Yoshitake2016}, and continues to decrease around $T_L$ reaching to almost zero below $T\sim 0.005$. 
The low-$T$ suppression is due to a nonzero flux gap required to excite the $Z_2$ gauge fluxes from the flux-free ground state~\cite{Kitaev2006}. 

\begin{figure}[htp]
    \includegraphics[width=.88\columnwidth,clip]{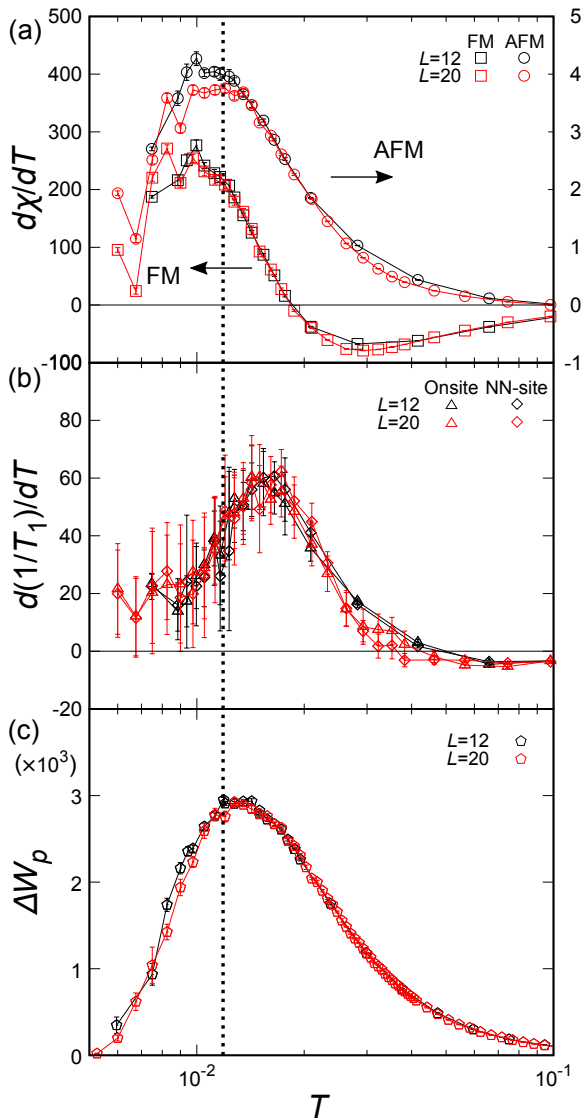}
    \caption{ \label{fig:fig3} 
    (a) and (b) $T$ derivatives of the data in Fig.~\ref{fig:fig2}. 
    (c) plots the thermal fluctuation of gauge fluxes $W_p$, $\Delta W_p$ in Eq.~(\ref{eq:Delta_W}). 
    The vertical dotted line indicates $T_L$. 
    }
\end{figure}

We also compute the $T$ derivatives of $\chi$ and $1/T_1$, as shown in Figs.~\ref{fig:fig3}(a) and \ref{fig:fig3}(b), respectively. 
Both derivatives show a peak around $T_L$, but change smoothly without showing any singularity. 
For comparison, we also compute the thermal fluctuation of gauge fluxes $W_p$ by the QMC method, defined by
\begin{align}
\Delta W_p = \frac{1}{N_p T^2}\Big( \big\langle \big(\sum_p W_p \big)^2 \big\rangle - \big\langle \sum_p W_p \big\rangle^2 \Big), 
\label{eq:Delta_W}
\end{align}
where $N_p$ is the number of plaquettes in the system. 
Note that $\Delta W_p$ corresponds to the specific heat in the anisotropic limit (toric code), where the effective Hamiltonian is given in the form $\mathcal{H} \propto \sum_p W_p$~\cite{Kitaev2006}; hence, $\Delta W_p$ measures the energy fluctuation related to the gauge fluxes. 
As shown in Fig.~\ref{fig:fig3}(c), $\Delta W_p$ also shows a broad peak around $T_L$, similar to the $T$ derivatives of $\chi$ and $1/T_1$. 
All these smooth changes with broad peaks are consistent with the fact that $T_L$ is not a phase transition but just a crossover in the 2D case~\cite{Nasu2015}. 
Furthermore, the similar behavior between three quantities in Fig.~\ref{fig:fig3} suggests that the $T$ derivatives of $\chi$ and $1/T_1$ provide good probes for the fluctuations of gauge fluxes. 

Interestingly, $d\chi/dT$ behaves differently between the FM and AFM cases, as shown in Fig.~\ref{fig:fig3}(a): 
it is negative for $T_L\lesssim T\lesssim T_H$ and changes the sign to positive just above $T_L$ for the FM case, while mostly positive in the same $T$ range for the AFM case. 
The qualitative difference will be useful for identifying the sign of the dominant Kitaev interactions in candidate materials. 
The details of the difference between the FM and AFM cases, including the nonlinear components of the magnetic susceptibility, will be reported elsewhere.

\subsection{3D hyperhoneycomb}
\label{subsec:3D}

\begin{figure}[htp]
    \includegraphics[width=.88\columnwidth,clip]{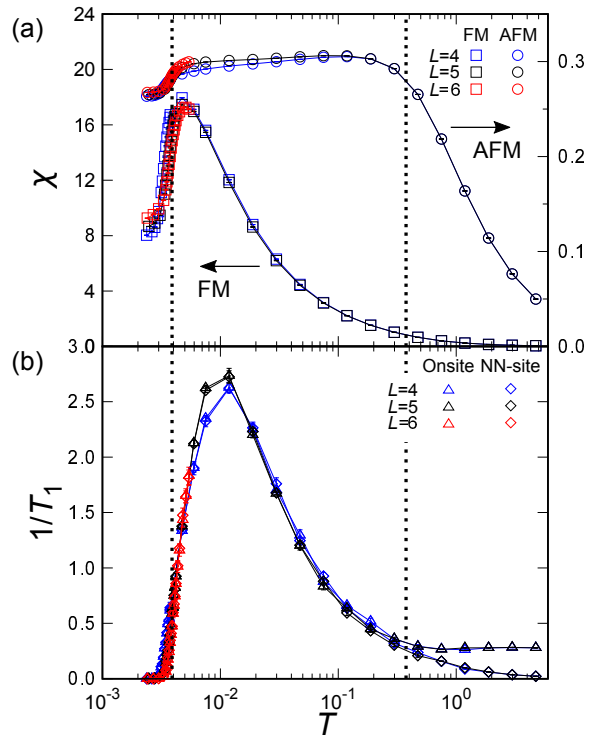}
    \caption{ \label{fig:fig4} 
    QMC+CTQMC results for the 3D hyperhoneycomb Kitaev model with isotropic $J_p$: (a) the magnetic susceptibility $\chi$ and (b) the NMR relaxation rate $1/T_1$. 
    The vertical dotted lines indicate $T_c\simeq 0.014$ and $T_H\simeq 0.375$~\cite{Nasu2014b}.
    Other notations are the same as those in Fig.~\ref{fig:fig2}. 
    }
\end{figure}

Next, we turn to the 3D case on the hyperhoneycomb lattice. 
Figure~\ref{fig:fig4} shows the QMC+CTQMC results for $\chi$ and $1/T_1$. 
The system size is given by $N=4L^3$: $256$, $500$, and $864$ sites for $L=4$, $5$, and $6$, respectively. 
Note that in the hyperhoneycomb lattice the $z$ bond is not equivalent to the $x$ and $y$ bonds from the lattice symmetry; we compute $\chi$ by Eq.~(\ref{eq:chi}) and $1/T_1$ by Eqs.~(\ref{eq:1/T_1}) and (\ref{eq:1/T_1_nnsite}) with replacing $\langle S^p_j (\tau) S^p_{j'}\rangle$ ($p=x,y$) by $\langle S^z_j (\tau) S^z_{j'}\rangle$ for simplicity. 
The overall $T$ dependence is similar to the 2D results as follows. 
The high-$T$ behaviors above $T_H$ are almost unchanged from the 2D cases, presumably because the bandwidth of matter fermions is independent of the dimensionality. 
With a decrease of $T$, $\chi$ begins to deviate from the Curie-Weiss behavior below $T\sim T_H$ and converges to a nonzero value after showing a peak, while $1/T_1$ increases below $T_H$ and strongly suppressed due to the flux gap after showing a peak at a low $T$. 
Nonetheless, there are quantitative differences. For instance, the peak of $\chi$ for the FM case is more than twice larger that that for the 2D case. 
Simultaneously, the change at low $T$ is much steeper in 3D than 2D. 
Similar behaviors are also seen in $1/T_1$. 
We will briefly comment on the quantitative differences in the end of this section. 

\begin{figure}[htp]
    \includegraphics[width=.88\columnwidth,clip]{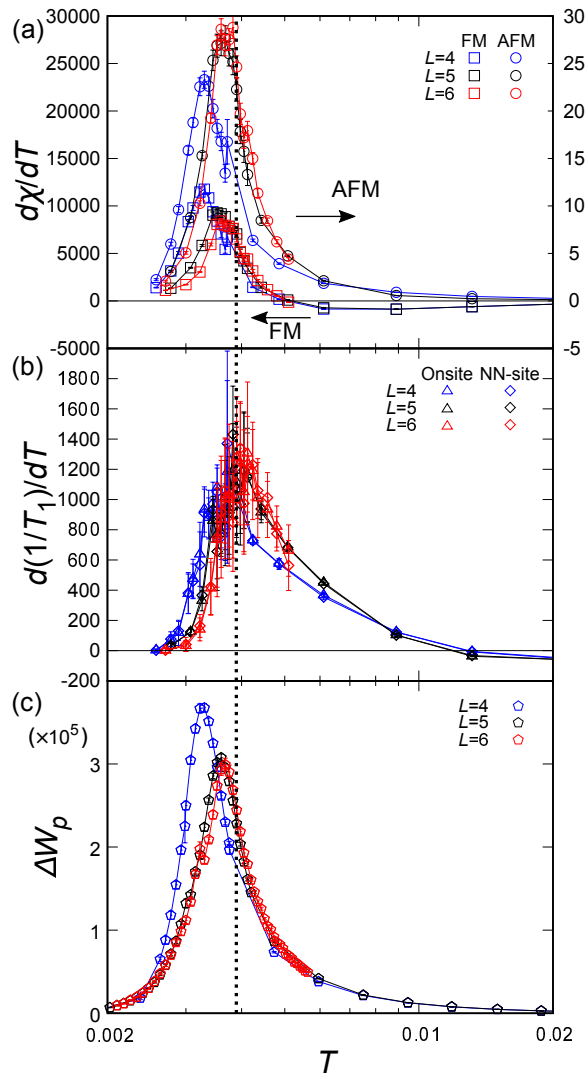}
    \caption{ \label{fig:fig5} 
    (a) and (b) $T$ derivatives of the data in Fig.~\ref{fig:fig4}. 
    (c) plots the thermal fluctuation of gauge fluxes $W_p$, $\Delta W_p$ in Eq.~(\ref{eq:Delta_W}). 
    The vertical dotted line indicates $T_c$. 
    }
\end{figure}

However, we also find a qualitative difference between 3D and 2D in the low-$T$ behavior. 
The 3D hyperhoneycomb model exhibits a phase transition at $T_c \simeq 0.0039$~\cite{Nasu2014b}. 
The phase transition takes place between the high-$T$ paramagnet and the low-$T$ QSL, driven by the proliferation of loops composed of the localized $Z_2$ gauge fluxes $W_p$. 
Thus, the transition is of topological nature, not characterized by local spin operators contrary to conventional magnetic ordering~\cite{Nasu2014b,Nasu2014}. 
Nevertheless, we find singular behaviors in both $\chi$ and $1/T_1$, as more clearly seen in the $T$ derivatives shown in Figs.~\ref{fig:fig5}(a) and \ref{fig:fig5}(b). 
Both $T$ derivatives show a sharp peak at $T\simeq T_c$, which becomes sharper for larger system sizes. 
We also plot the thermal fluctuation of gauge fluxes $\Delta W_p$ in Eq.~(\ref{eq:Delta_W}) in Fig.~\ref{fig:fig5}(c). 
In this 3D case, $\Delta W_p$ shows a similar sharp peak to $d\chi/dT$ and $d(1/T_1)/dT$. 
All these behaviors are in stark contrast to the 2D case, where the crossover at $T_L$ leads to smooth $T$ dependence as shown in Fig.~\ref{fig:fig3}. 

The low-$T$ behaviors of the dynamical quantities are substantially different from those in 2D, not only in the critical behavior associated with the phase transition but also the larger $T$ dependence. 
This clear difference depending on the spatial dimension is rather surprising when considering that the static spin correlations are not much different between 2D and 3D in the whole $T$ range~\cite{Nasu2015,Nasu2014b}. 
In the previous CDMFT+CTQMC study, the authors unveiled a prominent feature of the Kitaev QSL, the dichotomy between static and dynamical spin correlations, from the $T$ dependence of the static spin correlations for NN sites and $1/T_1$~\cite{Yoshitake2016,Yoshitake_preprint}. 
The significant dimensional dependence at low $T$ found here is another aspect of the dichotomy. 

We note that the difference of the sign of $d\chi/dT$ between the FM and AFM cases for $T_c \lesssim T \lesssim T_H$ is also seen in the 3D case, as shown in Fig.~\ref{fig:fig3}(c). 
We also note that the behavior of $d\chi/dT$ is similar to that found in the effective model in the anisotropic limit $J_z \gg J_x, J_y$~\cite{Nasu2014}.

\begin{figure*}[htp]
    \includegraphics[width=2\columnwidth,clip]{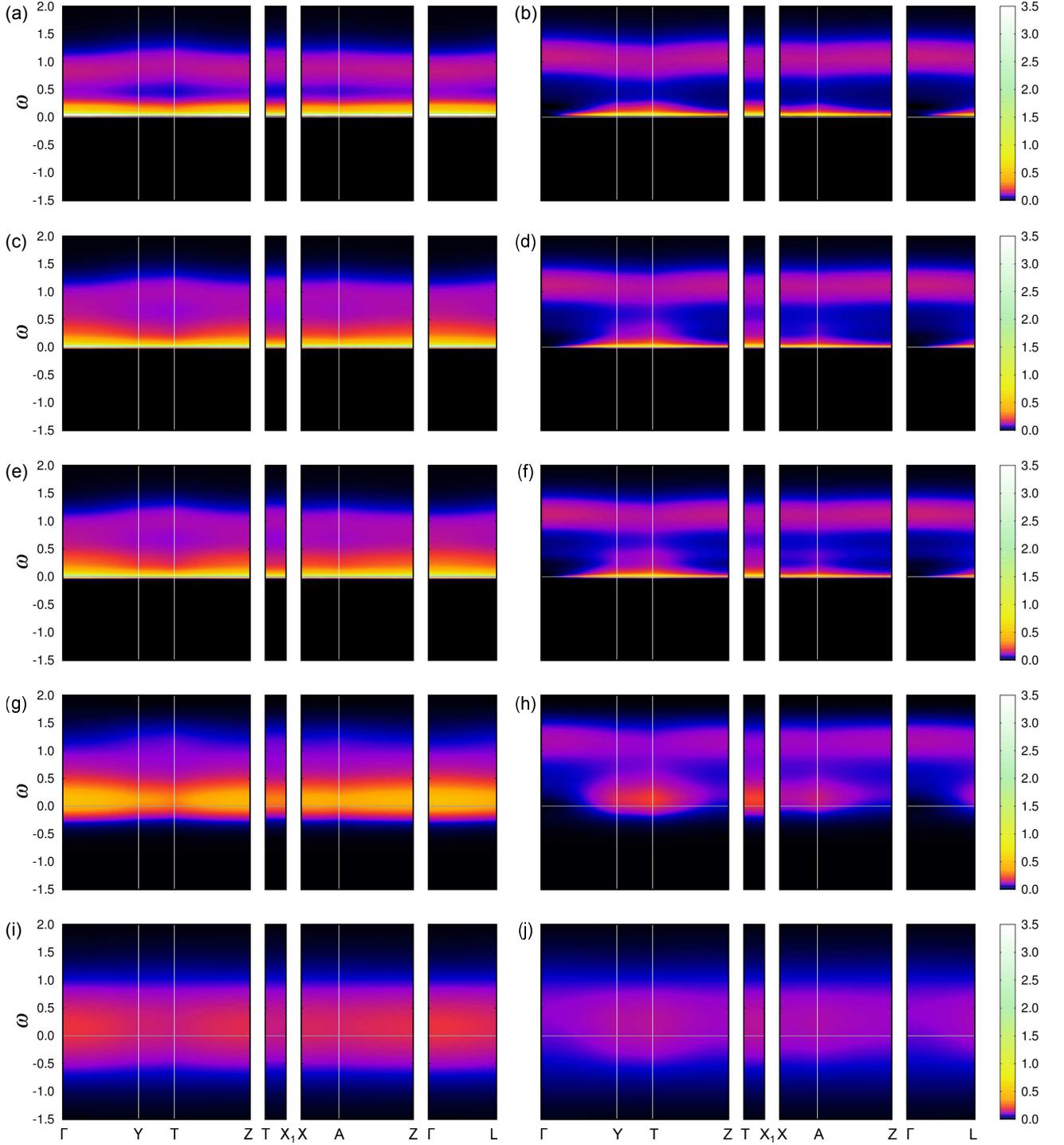}
    \caption{ \label{fig:fig6} 
    QMC+CTQMC results of the dynamical spin structure factor for the 3D hyperhoneycomb Kitaev model with isotropic $J_p$. 
    The data are calculated for $L=5$ ($500$ sites) and plotted along the symmetric lines indicated in Fig.~\ref{fig:fig1}(c). 
    (a)(c)(e)(g)(i) are for the FM case and (b)(d)(f)(h)(j) are for the AFM case: (a)(b) $T=0.002475$, (c)(d) $T=0.005955$, (e)(f) $T = 0.01185$, (g)(h) $T = 0.18825$, and (i)(j) $T = 1.185$. 
    Note that $T_c\simeq 0.0039$ and $T_H\simeq 0.375$~\cite{Nasu2014b}.
    }
\end{figure*}

Finally, we show the QMC+CTQMC results for the dynamical spin structure factor $S(\mathbf{q},\omega)$ for the 3D case in Fig.~\ref{fig:fig6}. 
$S(\mathbf{q},\omega)$ is defined as 
\begin{align}
S(\textbf{q}, \omega) = \frac1N \sum_{j,j'} e^{i\textbf{q}\cdot(\textbf{r}_{j}-\textbf{r}_{j'})}S^z_{j,j'}(\omega), 
\label{eq:S^p(q,w)}
\end{align}
where $\mathbf{r}_j$ represents the position vector for site $j$. 
The results are plotted along the symmetric lines in the first Brillouin zone shown in Fig.~\ref{fig:fig1}(c). 
The overall $T$ and $\omega$ dependence is similar to the 2D case reported in the previous study~\cite{Yoshitake2016}: 
almost $\mathbf{q}$-independent incoherent response around $\omega=0$ for $T\gtrsim T_H$, growth of the incoherent spectra around $\omega=|J|$ below $T_H$, and a rapid increase of the quasi-elastic response while approaching $T_c$. 
Also, as in the 2D case, the difference in the sign of $J$ appears in the $\mathbf{q}$  dependence of the spectral intensity. 
We note that the lowest-$T$ data below $T_c$ in Fig.~\ref{fig:fig4}(a) agree well with the previous $T=0$ result~\cite{Smith2015}.

\begin{figure*}[htp]
    \includegraphics[width=2\columnwidth,clip]{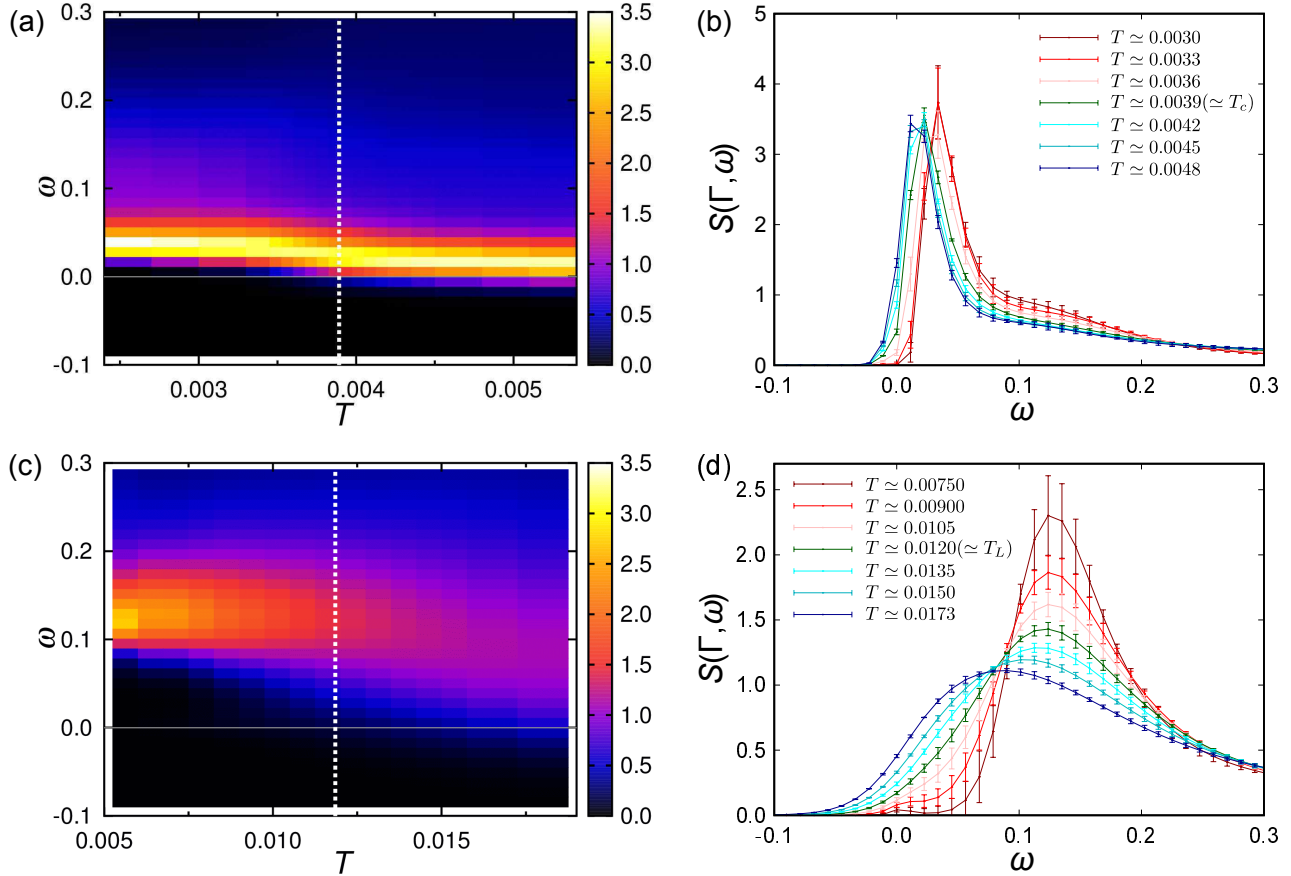}
    \caption{ \label{fig:fig7} 
    Comparison of the low-$\omega$ behaviors of $S(\Gamma,\omega) = S(\mathbf{q}=0,\omega)$ between the 3D and 2D cases at low $T$. 
    (a) Contour plot of $S(\Gamma,\omega)$ around $T_c$ for the 3D case and (b) the $\omega$ profiles. 
    (c) and (d) display the corresponding 2D results around $T_L$. 
    The data for the 3D and 2D cases are calculated for $L=6$ ($864$ sites) and $L=20$ ($800$ sites), respectively. 
    The white dotted lines in (a) and (c) indicate $T_c$ and $T_L$, respectively. 
    }
\end{figure*}

Figures~\ref{fig:fig7}(a) and \ref{fig:fig7}(b) display the low-$\omega$ part of $S(\Gamma,\omega) = S(\mathbf{q}=0,\omega)$ around $T_c$ for the FM case. 
Qualitatively similar behaviors are also seen for $S(\mathbf{q},\omega)$ near the zone boundary for the AFM case. 
With a decrease of $T$ across $T_c$, the quasi-elastic peak near $\omega= 0$ shifts to a slightly higher $\omega$, leading to the opening of the flux gap below $T_c$. 
The peak height is almost unchanged across $T_c$. 
For comparison, we plot the corresponding data for the 2D honeycomb case around $T_L$ in Figs.~\ref{fig:fig7}(c) and \ref{fig:fig7}(d). 
In the 2D case, the peak above $T_L$ is much broader with a lower peak height compared to the 3D case. 
When lowering $T$ across $T_L$, the peak becomes sharper with a shift of the peak position to a higher $\omega$. 
These differences between 2D and 3D are closely related with the quantitatively different behaviors of $\chi$ and $1/T_1$ observed in Figs.~\ref{fig:fig2} and \ref{fig:fig4} as follows. 
The sharper peak of $S(\Gamma,\omega)$ near $\omega= 0$ already existing above $T_c$ in 3D corresponds to much larger values of $\chi$ and $1/T_1$ just above $T_c$ compared to the 2D results above $T_L$. 
Furthermore, the shift of the peak across $T_c$ in Fig.~\ref{fig:fig7}(b) is related with the steep changes of $\chi$ and $1/T_1$ around $T_c$.

\section{Summary}
\label{sec:summary}

We have developed the numerical method for studying the spin dynamics of the Kitaev models by combining the QMC and CTQMC methods on the basis of a Majorana fermion representation. 
The QMC+CTQMC method overcomes the shortcoming in the previous CDMFT+CTQMC method, and enables us to investigate the very low-$T$ region where the $Z_2$ gauge fluxes play a role. 
The experimental observation of the gauge fluxes is one of the open issues in the Kitaev-type QSLs, and hence, the theoretical results obtained by our method provide the references for the experiments in candidate materials. 

We have applied the QMC+CTQMC method to the 2D and 3D Kitaev models. 
Calculating the magnetic susceptibility, the NMR relaxation rate, and the dynamical spin structure factor, we discussed the influences of thermally fluctuating gauge fluxes, with focusing on the differences arising from the spatial dimensions. 
In the 2D honeycomb case, everything changes smoothly while lowering $T$, reflecting the crossover associated with particlelike gauge flux excitations. 
In contrast, in the 3D hyperhoneycomb case, the system exhibits a phase transition by the proliferation of looplike gauge flux excitations, which leads to singular behaviors in the dynamical properties. 
We found that the dichotomy between static and dynamical spin correlations, which begins below the high-$T$ crossover associated with itinerant matter fermions, persists down to the low-$T$ region, in a more peculiar form reflecting thermally excited gauge fluxes; while the dichotomy in the higher-$T$ region is rather universal independent of the spatial dimension, the low-$T$ one appears differently between 2D and 3D, reflecting the different nature of the localized $Z_2$ gauge flux excitations. 
We showed that the $T$ derivatives of the magnetic susceptibility and the NMR relaxation rate provide good probes for fluctuating gauge fluxes in both 2D and 3D. 
Our results will be useful for identifying the contributions from the $Z_2$ gauge fluxes in the experimental candidates and also the nature of their excitations. 

\begin{acknowledgments}

This research was supported by Grants-in-Aid for Scientific Research under Grants No.~JP15K13533, No.~JP16K17747, No.~JP16H02206, and No.~JP16H00987. 
Parts of the numerical calculations were performed in the supercomputing systems in ISSP, the University of Tokyo. 

\end{acknowledgments}

\end{document}